\documentclass[preprint,showkeys,preprintnumbers,amsmath,amssymb,superscriptaddress]{revtex4}

\usepackage{natbib}
\usepackage{graphicx}

\begin{document}

\title{Entropy and complexity analysis of the $D$-dimensional rigid rotator and hyperspherical harmonics}

\author{J.S. Dehesa}
\email{dehesa@ugr.es}
\affiliation{Instituto Carlos I de F\'isica Te\'orica y Computacional, Universidad de Granada, Granada, Spain}
\affiliation{Departamento de F\'isica At\'omica, Molecular y Nuclear, Universidad de Granada, Granada, Spain}

\author{A. Guerrero}
\email{agmartinez@ugr.es}
\affiliation{Instituto Carlos I de F\'isica Te\'orica y Computacional, Universidad de Granada, Granada, Spain}
\affiliation{Departamento de F\'isica At\'omica, Molecular y Nuclear, Universidad de Granada, Granada, Spain}

\author{P. S\'anchez-Moreno}
\email{pablos@ugr.es}
\affiliation{Instituto Carlos I de F\'isica Te\'orica y Computacional, Universidad de Granada, Granada, Spain}
\affiliation{Departamento de Matem\'atica Aplicada, Universidad de Granada, Granada, Spain}

\begin{abstract}
In this paper we carry out an information-theoretic analysis of the $D$-dimensional rigid rotator by studying the entropy and complexity
measures of its wavefunctions, which are controlled by the hyperspherical harmonics. These measures quantify single and two-fold facets
of the rich intrinsic structure of the system which are manifest by the intricate and complex variety of D-dimensional geometries of the 
hyperspherical harmonics. We calculate the explicit expressions of the entropic moments and the R\'enyi entropies as well as the Fisher-R\'enyi,
Fisher-Shannon and LMC  complexities of the system. The explicit expression for the last two complexity measures is not yet possible, mainly because 
the logarithmic functional of the Shannon entropy has not yet been obtained up until now in a closed form.
\end{abstract}

\keywords{Hyperspherical harmonics, rigid rotor, complexity measures, Shannon entropy, R\'enyi entropy, Fisher information}

\maketitle

\section{Introduction}

The manifestations of quantum mechanics in $D$-dimensional physical systems are generally analytically inaccessible, basically because the associated Schr\"odinger equations cannot be explicitly solved except for a very few cases which correspond to a quantum potential with some known symmetry.
The particle-in-a-box, the harmonic oscillator, the hydrogen atom, the particle moving in a Dirac-delta-like potential, and the rigid rotator are possibly the five major prototypical systems which are used to model the quantum-mechanical behavior of most $3$- and $D$-dimensional physical systems (see e.g. \cite{galindo_1,herschbach_1}).

The information-theoretic properties of these physical prototypes have been recently investigated for the first four cases in references \cite{bouvrie_1,dehesa_3,dehesa_5,dehesa_6,lopez_1}; see also the review papers \cite{dehesa_4,dehesa_7}. However, the corresponding properties for the rigid rotator have not yet been found, although many other properties of this system are well known, such as the specific heat \cite{curilef_1}, potential energy surfaces \cite{rist_jmc12}, spectral quantities in external fields \cite{gartner_arxiv13}, among others. This is a serious lack because of the numerous applications of this model; in particular, it has been extensively used to characterize the rotation of diatomic molecules (and is easily extended to linear polyatomic molecules). In this work we investigate the entropy and complexity properties of the wavefunctions of the rigid rotator; i.e., the hyperspherical harmonics.

The $D$-dimensional ($D\geq 3$)  spherical harmonics (or simply, hyperspherical harmonics) do not only play a central role in harmonic analysis and approximation theory \cite{dai_13,muller_1,atkinson12} but also in quantum theory \cite{avery_3,avery_2}. As well, they have been shown to be the solutions of a very broad
class of equations of a form into which numerous equations of $D$-dimensional physics can be transformed, ranging from the Schr\"odinger 
equation of the rigid rotator till the Bethe-Salpeter equation of some quark systems 
\cite{muller_1,kyriakopoulos_1,stein_1,aquilanti_1,avery_3,nikiforov_2,herschbach_1,avery_4,avery_2}.
Indeed, e.g. they are the eigenfunctions of the $D$-dimensional rigid rotator (i.e., a point mass $\mu$ rotating around a fixed center in the hyperspace 
at a given distance $r_0$) corresponding to the eigenvalues $l(l+D-2)/(2I)$, for $l = 0,1,2,\ldots$, where the moment of inertia $I =\mu r_{0}^2$. 
Moreover, they are the functions that give the anisotropic character of the eigenfunctions of $D$-dimensional central potentials, since the remaining 
radial part is spherically symmetric. The hyperspherical harmonics are functions defined on the ($D-1$)-dimensional unit sphere 
$S_{D-1} \subset \mathbb{R}_D$ which arise as eigenfunctions of the Laplace-Beltrami operator corresponding to the eigenvalues $l(l+D-2)$. They
are basis vectors in certain irreducible representation spaces of $SO(D, 2)$ \cite{dai_13,muller_1,atkinson12}, and in fact constitute a basis for integrable functions defined 
on the unit sphere. 

The hyperspherical harmonics are known \cite{avery_3,nikiforov_2,dehesa_3}
to have the form

\begin{equation}
   \label{hyperspherical_harmonics}
   Y_{l,\{\mu\}}(\Omega_D)=\frac{1}{\sqrt{2\pi}} e^{i\mu_{D-1}\theta_{D-1}} \prod_{j=1}^{D-2}   \hat{C}_{\mu_j-\mu_{j+1}}^{\alpha_j+\mu_{j+1}} (\cos \theta_j) (\sin \theta_j)^{\mu_{j+1}},
\end{equation}
where $\Omega_D \equiv (\theta_1,\theta_2,\cdots,\theta_{D-1})$ represents the $D-1$ angular coordinates of the sphere $S_{D-1}$ so that
$0\leq \theta_j \leq \pi$ for $j=1,\cdots,D-2$ and $0 \leq \theta_{D-1} \leq 2\pi$. The $D-1$ integer numbers $l \equiv \mu_1$ and
$\{\mu_2,\cdots,\mu_{D-1}\equiv m\} \equiv \{\mu\}$ have the values $l=0,1,2,\cdots$ and $\mu_1\geq \mu_2 \geq \ldots \geq |\mu_{D-1}|\ge 0$.
The parameter $\alpha_j=(D-j-1)/2$. And the symbol $\hat{C}^{\lambda}_n(x)$, $\lambda>-\frac{1}{2}$, denotes the orthonormal Gegenbauer polynomial of 
degree $n$ and parameter $\lambda$ which satisfies the orthogonality condition

\begin{equation}
   \label{orthonormalization_gegenbauer}
   \int_{-1}^{+1} \hat{C}^{\lambda}_n(x) \hat{C}^{\lambda}_m(x) \omega_{\lambda}(x) dx=\delta_{mn},
\end{equation}
where the weight function is defined as
\begin{equation}
\omega_{\lambda}(x)=(1-x^2)^{\lambda-\frac{1}{2}}.
\label{eq:weight_gegenbauer}
\end{equation}

The algebraic properties of these functions are widely
known in mathematical physics \cite{wen_1,aquilanti_1,avery_3,nikiforov_2,herschbach_1,avery_4,avery_1,avery_2};
in particular, they satisfy the orthogonality relation
\[
  \int_{S_{D-1}} Y_{l,\{\mu\}}^*(\Omega_D) Y_{l',\{\mu'\}}(\Omega_D) d\Omega_D=\delta_{ll'} \delta_{\{\mu\},\{\mu'\}},
\]
where the generalized solid angle element is
\[
d\Omega_D=\left(\prod_{j=1}^{D-2} (\sin \theta_j)^{2 \alpha_j} d\theta_j\right) d\theta_{D-1}.
\]
The spread of the hyperspherical harmonics all over the hyperspace is, however, not so well known. This is a serious lack since these functions
control the angular distribution of the charge and momentum distributions of numerous quantum mechanical systems with a central potential, by means of the density function
\begin{equation}
\label{density_harmonics}
  \rho_{l,\{\mu\}}(\Omega_D)=\left|Y_{l,\{\mu\}}\left(\Omega_D\right)\right|^2,
\end{equation}
which is called as Rakhmanov probability density of the hyperspherical harmonics in the theory of special functions, and gives
the distribution of the particle all over the hyperspace. The information-theoretic measures of this density function allows us to quantify
single and composite facets of the rich variety of $D$-dimensional geometries of the system in the hyperspace.

The goal of this paper is three-fold. First, we calculate the analytical expressions of various single information-theoretic measures of
spreading (entropic moments and R\'enyi entropies) beyond the recently found Fisher information \cite{dehesa_2}, and the
following two-component complexity measures: Fisher-Shannon, Fisher-R\'enyi and LMC complexities. Second, we apply these results to
eigenfunctions of the standard (i.e., three-dimensional) rigid rotator; that is to the hyperspherical harmonics. Third, we carry out a numerical
study of these entropy and complexity quantities for various orders and dimensionalities of the harmonics.

The structure of the paper is the following. In Section \ref{sec:basics} we give the definitions of the entropies and complexities to be used throughout
the paper. Then, in Section \ref{sec:information_measures} we give the expression of the Fisher information and calculate the entropic moments and R\'enyi entropies of the wavefunctions of the quantum-mechanical
$D$-dimensional rigid rotator, which are controlled by the hyperspherical harmonics. In Section \ref{sec:complexity_measures} the expressions of the two-component
complexity measures of the type Fisher-Shannon, Fisher-R\'enyi and LMC types are given, and a numerical study is performed.
Finally, some conclusions are given and various open problems are pointed out.

\section{Entropy and complexity measures: Basics}
\label{sec:basics}

In this Section we describe briefly the information-theoretic spreading measures of a general probability density $\rho(\vec{{\bf r}})$
which will be used throughout the paper; namely, the entropic moments, the R\'enyi, Tsallis and Shannon entropies, the Fisher information and the
associated two-component complexity measures: Fisher-Shannon, Fisher-R\'enyi, and LMC.

The $q$th-frequency or entropic moment of the density $\rho(\vec{{\bf r}})$, $\vec{{\bf r}} \in \mathbb{R}^D$, is defined by
\begin{equation}
\label{entropic_moment_q}
  W_q[\rho]:=\langle \rho^{q-1} \rangle=\int_{\mathbb{R}^D} \left[\rho(\vec{{\bf r}})\right]^q d \vec{{\bf r}}, \quad q \in \mathbb{R}^+
\end{equation}
where the expectation value of a function $f(\vec{{\bf r}})$, $\langle f(\vec{{\bf r}}) \rangle$, is given by
\[
 \langle f(\vec{{\bf r}}) \rangle = \int_{\mathbb{R}^D} f(\vec{{\bf r}}) \rho(\vec{{\bf r}}) d\vec{{\bf r}}.
\]
Mathematically, these moments are often more useful than the ordinary moments $\langle r^k \rangle$ because the later ones give too much
weight to the tail of the distribution and, at times, they are undefined \cite{uffink_1}. From a physical point of view the entropic 
moments describe numerous functionals of the electron density which characterize fundamental and/or experimentally-measurable
quantities of atomic and molecular systems according to the Hohenberg-Kohn density-functional theory \cite{parr_1,lin_1,lin_2,nagy_1}; e.g. the
Thomas-Fermi and Dirac exchange energies. See also \cite{angulo_2} for their connection with other atomic
density functionals, \cite{pinta_1,romera_1} for the existence conditions, \cite{leo_1} for further mathematical properties, \cite{dehesa_4} for various applications in $D$-dimensional quantum systems, and \cite{dette_as05} for potential applications in statistics and imaging.

The R\'enyi and Tsallis entropies of $\rho(\vec{{\bf r}})$ are defined in terms of the entropic moments as \cite{renyi_1}
\begin{equation}
\label{renyi_entropy}
  R_q[\rho]=\frac{1}{1-q} \log W_q[\rho]=\frac{1}{1-q} \log \int_{\mathbb{R}^D} \left[\rho(\vec{{\bf r}})\right]^q d\vec{{\bf r}}, \quad
  q>0, \,\,q \neq 1,
\end{equation}
and \cite{tsallis_jsp88}
\begin{equation}
\label{tsallis_entropy}
  T_q[\rho]=\frac{1}{q-1} \left(1 - W_q[\rho]\right)=\frac{1}{q-1} \left(1- \int_{\mathbb{R}^D} \left[\rho(\vec{{\bf r}})\right]^q d\vec{{\bf r}}\right), \quad
  q>0, \,\,q \neq 1,
\end{equation}
respectively, which when $q \rightarrow 1$ reduce to the well-known Shannon entropy
\begin{equation}
\label{shannon_entropy}
S[\rho]=-\int_{\mathbb{R}^D}\rho(\vec{{\bf r}}) \log \rho(\vec{{\bf r}}) d\vec{{\bf r}}.
\end{equation}

It is interesting to remark that these quantities are global measures of spreading of the density $\rho(\vec{{\bf r}})$ because they are power
(R\'enyi) or logarithmic (Shannon) functionals of $\rho(\vec{{\bf r}})$. They provide various complementary ways to quantify the extent of $\rho(\vec{{\bf r}})$
all over the hyperspace.

The (translationally invariant) Fisher information of $\rho(\vec{{\bf r}})$ is defined \cite{fisher_1,frieden_1} by

\begin{equation}
 \label{fisher_information}
  F[\rho]=\int_{\mathbb{R^D}} \rho(\vec{{\bf r}}) \left| \nabla_D \log \rho(\vec{{\bf r}})\right|^2 d\vec{{\bf r}}=
  4 \int_{\mathbb{R^D}} \left| \nabla_D \sqrt{\rho(\vec{{\bf r}})}\right|^2 d\vec{{\bf r}},
\end{equation}
where $\nabla_D$ denotes the $D$-dimensional gradient. This notion was first introduced in the one-dimensional case for statistical estimation 
\cite{fisher_1}, but nowadays it is used in a wide variety of scientific fields \cite{frieden_1} mainly because of its close resemblance with 
kinetic and Weisz\"acker energies \cite{sears_1}. Contrary to the R\'enyi and Shannon entropies, the Fisher information is a local measure of 
spreading of the density because it is a gradient functional of $\rho(\vec{{\bf r}})$. The higher this quantity is, the more localized is the
density, the smaller is the uncertainty and the higher is the accuracy in estimating the localization of the particle.

Recently, some composite density-dependent information-theoretic quantities have been introduced. They are called complexity measures
because they grasp more than a single facet (macroscopic property) of the density. We refer to the Fisher-Shannon, and the more general Fisher-R\'enyi, and the LMC shape complexities.
They have a number of very interesting
mathematical properties. Here we would like to highlight some common characteristics. They are
dimensionless,
opposite to the previously defined single-component entropies (entropic moments, Shannon and R\'enyi entropies, Fisher information), what allows
them to be mutually compared. They are defined essentially by the product of two single entropies, what allows them to quantify two-fold
facets of the density. Moreover, they are intrinsic quantities of the density what differenciate them from other complexity notions already
used (computational complexity, algorithmic complexity, ...), which depend on the context. Finally, they are close to the intuitive notion
of complexity because they are minimum for the extreme or least complex distribution which correspond to maximum disorder (i.e. the highly flat distribution).

The Fisher-R\'enyi complexity of $\rho(\vec{{\bf r}})$ is defined \cite{romera_2} by
\begin{equation}
 \label{fisher_renyi}
C_{FR}^{(q)}\left[\rho\right]:= F\left[\rho\right] \times J_q\left[\rho\right]
\end{equation}
where $F\left[\rho\right]$ is the Fisher information (\ref{fisher_information}) and $J_q\left[\rho\right]$ denotes the $q$th-order R\'enyi power entropy of 
$\rho(\vec{{\bf r}})$ given by
\begin{equation}
\label{renyi_power_entropy}
J_q\left[\rho\right]=\frac{1}{2 \pi e} e^{\frac{2}{D} R_q\left[\rho\right]}
\end{equation}
where $R_q\left[\rho\right]$ is the R\'enyi entropy (\ref{renyi_entropy}). This complexity measure quantifies wiggliness or gradient content of the density
jointly with its total extent all over the hyperspace, the parameter $q$ weighting different regions of $\rho(\vec{{\bf r}})$. The special
case $q \to 1$ of (\ref{fisher_renyi}) leads to the Fisher-Shannon complexity as
\begin{equation}
 \label{fisher_shannon_complexity}
C_{FS}\left[\rho\right]=F\left[\rho\right] \times \frac{1}{2 \pi e} e^{\frac{2}{D} S\left[\rho\right]},
\end{equation}
where $S\left[\rho\right]$ is the Shannon entropy (\ref{shannon_entropy}). All the relevant invariance properties (replication, translation, scaling) of 
$C_{FS}\left[\rho\right]$ are also fulfilled by the Fisher-R\'enyi complexities $C_{FR}^{(q)}\left[\rho\right]$ for any $q>0$, $q \neq 1$.

The LMC complexity of $\rho(\vec{{\bf r}})$ is given \cite{catalan_2,lopez_ruiz_2} by
\begin{equation}
 \label{lmc_complexity}
C_{LMC}\left[\rho\right]= D\left[\rho\right] \times e^{S\left[\rho\right]},
\end{equation}
where
\begin{equation}
D\left[\rho\right]=W_2\left[\rho\right]=\langle \rho \rangle
\label{eq:def_disequilibrium}
\end{equation}
is the second-order entropic moment of $\rho$,
also called disequilibrium in some contexts. This complexity measure quantifies the combined balance of the average height of
$\rho(\vec{{\bf r}})$ and the total extent of the spread of the density over the whole hyperspace.

\section{Entropy measures of hyperspherical harmonics}
\label{sec:information_measures}
In this Section we give the algebraic expression of the Fisher information and obtain those of the entropic moments and R\'enyi and Tsallis entropies of the hyperspherical harmonics
$Y_{l,\{\mu\}}(\Omega_D)$, which are given by the corresponding quantities, $F[\rho]$, $W_q\left[\rho\right]$ and $R_q\left[\rho\right]$ respectively,
of the associated Rakhmanov probability density $\rho=\rho_{l,\{\mu\}}(\Omega_D)$. They will be expressed in terms of the hyperquantum
numbers $\left(\mu_1\equiv l, \mu_2, \ldots, \mu_{D-1}\right)\equiv \left(l,\left\{\mu\right\}\right)$ and the dimensionality $D$.

First we realize from Eqs. (\ref{hyperspherical_harmonics}) and (\ref{density_harmonics}) that the Rakhmanov density of the 
hyperspherical harmonics is 
\begin{equation}
\label{density_harmonics_2}
\rho_{l,\{\mu\}}(\Omega_D)=\frac{1}{2\pi} \prod_{j=1}^{D-2} \left[\hat{C}_{\mu_j-\mu_{j+1}}^{\alpha_j+\mu_{j+1}} (\cos \theta_j)\right]^2 (\sin \theta_j)^{2\mu_{j+1}}.
\end{equation}

Then, according to Eq. (\ref{fisher_information}), the Fisher information of this density is \cite{romera_jmp06,dehesa_2}
\begin{equation}
           \label{fisher_dimension_d}
           F[\rho_{l,\{\mu\}}]=4L(L+1)-2|\mu_{D-1}|(2L+1)-(D-1)(D-3),
\end{equation}
where $L=l+\frac{D-3}{2}$. In the three-dimensional case ($D=3$) this yields
\begin{equation}
           \label{fisher_dimension_d3}
           F_{l,m}[\rho]=4l(l+1)-2|m|(2l+1).
\end{equation}

The entropic moments of this density are, according to Eq. (\ref{entropic_moment_q}),
\begin{eqnarray}
  \label{entropic_moment_harmonics}
W_q[\rho_{l,\{\mu\}}]&=&\int_{\mathbb{S}_{D-1}} \left[\rho_{l,\{\mu\}}(\Omega_D)\right]^q d \Omega_D \\ \nonumber
   &=& \frac{1}{(2\pi)^{q-1}} \prod_{j=1}^{D-2} \int_{0}^{\pi} \left|\hat{C}_{\mu_j-\mu_{j+1}}^{\alpha_j+\mu_{j+1}} (\cos \theta_j)\right|^{2q} 
      (\sin \theta_j)^{2(q\mu_{j+1}+\alpha_j)} d \theta_j
\end{eqnarray}

The change of variable $\theta_j \rightarrow x_j=\cos \theta_j$ allows us to write these quantities as follows
\begin{eqnarray}
  \label{entropic_moment_harmonics_2}
  W_q[\rho_{l,\{\mu\}}]&=& \frac{1}{(2\pi)^{q-1}} \prod_{j=1}^{D-2} \int_{-1}^{+1} \left|\hat{C}_{\mu_j-\mu_{j+1}}^{\alpha_j+\mu_{j+1}} (x_j)\right|^{2q}
                           \left(1-x_j^2\right)^{q\mu_{j+1}+\alpha_j-\frac{1}{2}} dx_j \\ 
   &=& \frac{1}{(2\pi)^{q-1}} \prod_{j=1}^{D-2} \int_{-1}^{+1} \left|\hat{C}_{\mu_j-\mu_{j+1}}^{\alpha_j+\mu_{j+1}} (x_j)\right|^{2q}
                           \omega_{q\mu_{j+1}+\alpha_j}(x_j)  \,dx_j
\end{eqnarray}
where $\omega_\lambda(x)$ is defined in (\ref{eq:weight_gegenbauer}).

For $q\in\mathbb{N}$ we can apply the linearization method for Jacobi polynomials by Srivastava \cite{srivastava_ass88}, particularized for Gegenbauer polynomials. This method yields the following linearization formula:
\[
 \left[\hat{C}_{\mu_j-\mu_{j+1}}^{\alpha_j+\mu_{j+1}} (x_j)\right]^{2q} = \sum_{i=0}^{2q(\mu_j-\mu_{j+1})} \beta_{j,q,D}^{(i)}
 \frac{d_i^{\left(q\mu_{j+1}+\alpha_j-\frac{1}{2},q\mu_{j+1}+\alpha_j-\frac{1}{2}\right)}}
 {\left[d_{\mu_j-\mu_{j+1}}^{\left(\mu_{j+1}+\alpha_j-\frac{1}{2},\mu_{j+1}+\alpha_j-\frac{1}{2}\right)}\right]^{2q}}
 \hat{C}_{i}^{\alpha_j+q\mu_{j+1}} (x_j)
\]
which, together with the orthogonality relation of the Gegenbauer polynomials, allows us to obtain the following expression for the entropic moments:
\begin{equation}
  \label{entropic__moment_harmonics_3}
    W_q[\rho_{l,\{\mu\}}]=\frac{1}{(2\pi)^{q-1}} \prod_{j=1}^{D-2} \beta_{j,q,D}^{(0)} 
    \frac{\left[d_0^{\left(q\mu_{j+1}+\alpha_j-\frac{1}{2},q\mu_{j+1}+\alpha_j-\frac{1}{2}\right)}\right]^{2}}{\left[d_{\mu_j-\mu_{j+1}}^{\left(\mu_{j+1}+\alpha_j-\frac{1}{2},\mu_{j+1}+\alpha_j-\frac{1}{2}\right)}\right]^{2q}},
\end{equation}
where
\begin{equation}
 \label{normalization_constant_jacobi}
  d_n^{(\alpha,\beta)}=\sqrt{\frac{2^{\alpha+\beta+1} \Gamma(n+\alpha+1) \Gamma(n+\beta+1)}{n! (2n+\alpha+\beta+1) \Gamma(n+\alpha+\beta+1)}}
\end{equation}
is the normalization constant of the Jacobi polynomials $P_n^{(\alpha,\beta)}(x)$ and 
\[
 \beta_{j,q,D}^{(0)}=c\left(2q,\mu_j-\mu_{j+1},\alpha_j+\mu_{j+1}-\frac{1}{2},\alpha_j+\mu_{j+1}-\frac{1}{2},\alpha_j+q\mu_{j+1}-\frac{1}{2},\alpha_j+q\mu_{j+1}-\frac{1}{2}\right)
\]
with
\begin{multline}
 \label{linearization_coefficient_jacobi}
    c(r,n,\alpha,\beta,\gamma,\delta)\\
    =
    \binom{n+\alpha}{n}^r
    F^{1:2;\ldots;2}_{1:1;\ldots;1}\left(
    \begin{array}{c}
    \gamma+1: -n,\alpha+\beta+n+1;\ldots;-n,\alpha+\beta+n+1\\
    \gamma+\delta+2:\alpha+1;\ldots;\alpha+1
    \end{array}
    ;1,\ldots,1
    \right)\\
    =
    \binom{n+\alpha}{n}^r 
    \sum_{j_1,\ldots,j_r=0}^n 
    \frac{(\gamma+1)_{j_1+\cdots+j_r}}{(\gamma+\delta+2)_{j_1+\cdots+j_r}}
    \frac{(-n)_{j_1}(\alpha+\beta+n+1)_{j_1}\cdots
    (-n)_{j_r}(\alpha+\beta+n+1)_{j_r}}{(\alpha+1)_{j_1}
    \cdots (\alpha+1)_{j_r} j_1!\cdots j_r!},
\end{multline}
where $F^{1:2;\ldots;2}_{1:1;\ldots;1}$ is a Srivastava-Daoust function \cite{srivastava_ass88}. This expression generalizes to any $q$ the expression of the entropic moment with $q = 4$ already obtained in \cite{dehesa_2}.

Let us now consider some examples: In the case $D=3$ we obtain the expressions
\[
W_q[\rho_{0,0}]=2^{2-2q}\pi^{1-q}
\]
for $l=m=0$,
\[
W_q[\rho_{1,0}]=\frac{2^{2-2q} 3^q \pi^{1-q}}{2q+1}
\]
for $l=1$, $m=0$,
\[
W_q[\rho_{l,l}]=(2\pi)^{1-q} \frac{2^{2ql+1}\left(\Gamma(ql+1)\right)^2}{(2ql+1)\Gamma(2ql+1)}
\left(
\frac{(2l+1)\Gamma(2l+1)}{2^{2l+1}\left(\Gamma(l+1)\right)^2}
\right)^q
\]
for $m=l$, and
\[
W_q[\rho_{l,l-1}]=(2\pi)^{1-q}
l^{2q}
\frac{\Gamma\left(q+\frac12\right) \Gamma\left(q(l-1)+\frac32\right)}{\sqrt{\pi}\Gamma\left(ql+\frac32\right)}
\frac{\left(d_0^{(q(l-1),q(l-1)}\right)^2}{\left(d_1^{(l-1,l-1}\right)^{2q}}
\]
for $m=l-1$.

For $D=2$ the spherical harmonic reduces to $Y_m(\theta)=\frac{1}{\sqrt{2\pi}}e^{im\theta}$, $m\in\mathbb{Z}$,  so the entropic moment of order $q$ have the constant value
\[
W_q[\rho_m]=(2\pi)^{1-q}.
\]

For $D=4$ we can obtain the values of the entropic moments
\[
W_q[\rho_{0,0,0}]=2^{1-q} \pi^{2-2q}
\]
for $\mu_1=\mu_2=\mu_3=0$,
\[
W_q[\rho_{1,0,0}]=\frac{2^{1+q} \pi^{\frac{3}{2}-2q} \Gamma\left(\frac{1}{2}+q\right)}{\Gamma(2+q)}
\]
for $\mu_1=1$ and $\mu_2=\mu_3=0$,
\[
W_q[\rho_{l,l,l}]=(2\pi^2)^{1-q} \frac{(l+1)^q}{lq+1}
\]
for $\mu_1=\mu_2=\mu_3=l$,
\[
W_q[\rho_{l,l-1,l-1}]=2 \pi^{\frac32-2q} (l(l+1))^q
\frac{\Gamma\left(q+\frac12\right)\Gamma\left(q(l-1)+1\right)}{\Gamma(lq+2)}
\]
for $\mu_l$ and $\mu_2=\mu_3=l-1$,
\[
W_q[\rho_{l-1,l-1,l-2}]=2^{1+q} \pi^{1-2q} (l(l^2+1))^q
\frac{\left(\Gamma\left(q+\frac12\right)\right)^2\Gamma\left(q(l-2)+1\right)}{\Gamma(lq+2)}
\]
for $\mu_l$, $\mu_2=l-1$ and $\mu_3=l-2$.

For any value of the dimensionality $D$ we can obtain the following results:
\[
W_q[\rho_{0,0,\ldots,0}]=(2\pi)^{1-q} 2^{(D-1)(D-2)(1-q)/2} ((D-2)!)^{q-1}
\prod_{j=1}^{D-2} \frac{\left(\Gamma\left(\frac{D-j}{2}\right)\right)^{2-2q}}{\left(\Gamma\left(D-j-1\right)\right)^{1-q}}
\]
for $\mu_1=\mu_2=\cdots=\mu_{D-1}=0$,
\begin{multline*}
W_q[\rho_{l,l,\ldots,l}]=(2\pi)^{1-q} 2^{(D-1)(D-2)(1-q)/2}
\frac{((2l+1)_{D-2})^q}{(2ql+1)_{D-2}}\\
\times \prod_{j=1}^{D-2} \frac{\left(\Gamma\left(ql+\frac{D-j}{2}\right)\right)^2}{\Gamma(2ql+D-j-1)}
\left(\frac{\Gamma(2l+D-j-1)}{\left(\Gamma\left(l+\frac{D-j}{2}\right)\right)^2}\right)^q
\end{multline*}
for $\mu_1=\mu_2=\cdots=\mu_{D-1}=l$.

These expressions together with Eqs. (\ref{renyi_entropy}) and (\ref{eq:def_disequilibrium}) allow us to obtain the R\'enyi and Tsallis entropies of the quantum-mechanical states of the $D$-dimensional rigid rotator, respectively, in a straightforward manner in terms of the hyperquantum numbers characterizing the states and the dimensionality $D$.

\section{Complexity measures of hyperspherical harmonics}
\label{sec:complexity_measures}

In this Section we consider the complexity measures of Fisher-Shannon, Fisher-R\'enyi and LMC of the eigenfunctions of the $D$-dimensional rigid rotator (i.e., the hyperspherical harmonics) which are described by the corresponding quantities of the associated probability density given by Eq. (\ref{density_harmonics}) or (\ref{density_harmonics_2}). We should immediately say that  these quantities cannot be obtained in analytical form, mainly because of the highbrow expression of the R\'enyi entropy (as seen in the previous section) and the logarithmic character of the Shannon functional. Therefore, our study has to be necessarily numerical. We will fix the dimensionality $D = 3$, so that we will investigate the behavior of the abovementioned complexity measures for the eigenfunctions of the three-dimensional rigid rotator (i.e., the standard spherical harmonics $Y_{l,m}(\theta,\phi)$) in terms of the quantum numbers $l$ and $m$.
We will numerically perform a complexity analysis of the three-dimensional rigid rotator (i.e. a point-mass particle freely moving on the two-dimensional sphere) whose ground and excited states $(l,m)$ have the associated probability density
\begin{equation}
\rho_{l,m}(\theta,\phi)=\frac{1}{2\pi} \left[\hat{C}_{l-m}^{\frac12+m} (\cos \theta)\right]^2 (\sin \theta)^{2m}.
\label{eq:3d_density}
\end{equation}
It is well known that this system models a great number of physical systems, such as e.g. the rotating diatomic molecules. Indeed, a diatomic molecule is an extremely complicated many body problem (e.g., the HCl molecule is a $20$-body problem), but at very low energies no excitations associated with the electron degrees of freedom come into play since the electron cloud binds the two atomic nuclei into a nearly rigid structure. For further details and applications of the three-dimensional rigid rotator, see e.g. \cite{muller_1,nikiforov_2,stein_1}.

\subsection{Fisher-Shannon complexity} 

\begin{figure}
\begin{center}
\includegraphics[width=10cm]{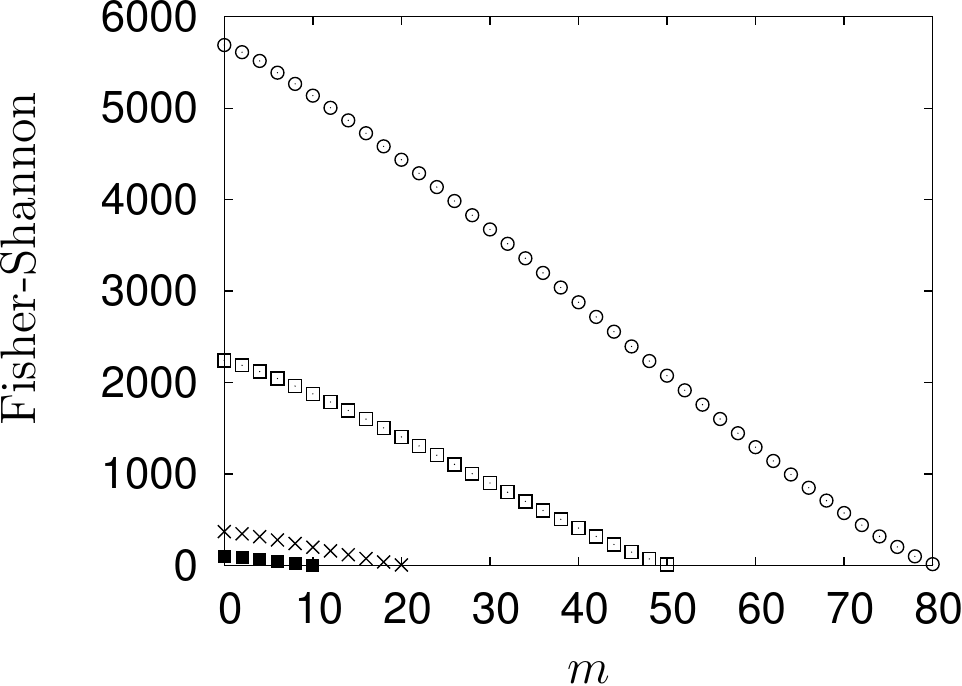}
\end{center}
\caption{Dependence of the Fisher-Shannon complexity on the magnetic quantum number $m = 0,\ldots, l$, for various spherical harmonics $Y_{l,m}(\theta, \phi)$ with a fixed orbital quantum number $l = 10$ ($\blacksquare$), $20$ ($\times$), $50$ ($\boxdot$) and $80$ ($\odot$).}
\label{fig_fs1}
\end{figure}

\begin{figure}
\begin{center}
\includegraphics[width=10cm]{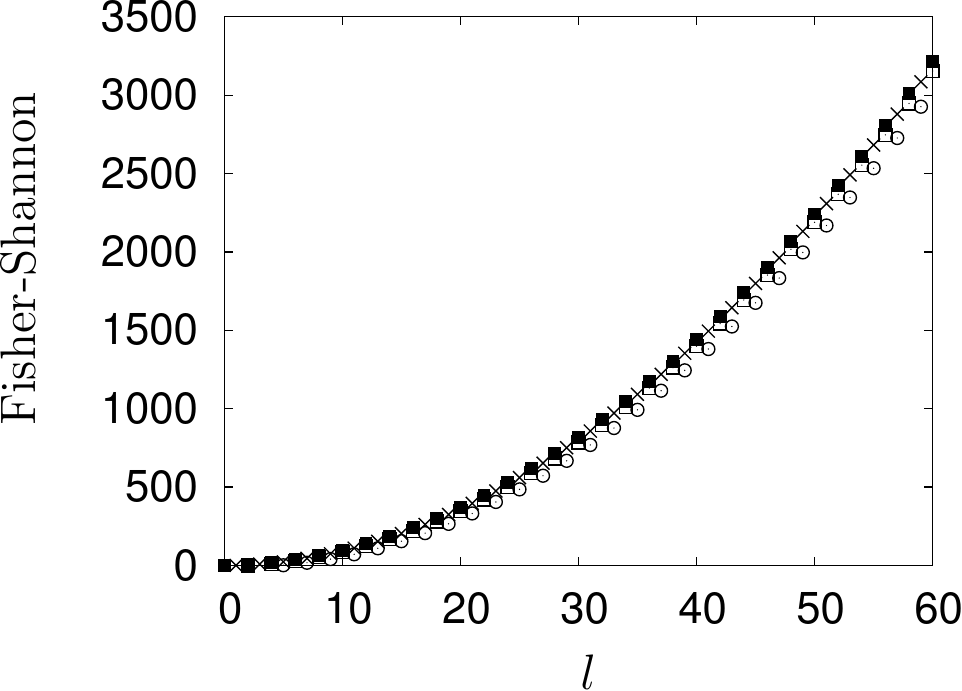}
\end{center}
\caption{Growth of the Fisher-Shannon complexity with $l$ for various spherical harmonics  $Y_{l,m}(\theta, \phi)$  for fixed $m = 0$ ($\blacksquare$), $1$ ($\times$), $2$ ($\boxdot$) and $5$ ($\odot$).}
\label{fig_fs2}
\end{figure}

\begin{figure}
\begin{center}
\includegraphics[width=10cm]{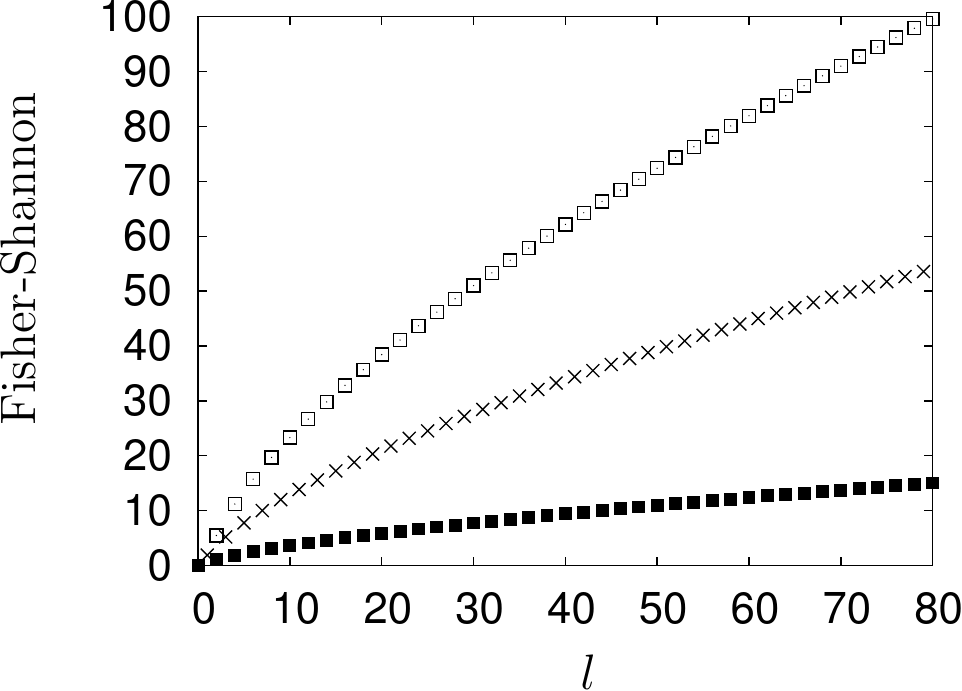}
\end{center}
\caption{Behaviour of the Fisher-Shannon complexity of the spherical harmonics $Y_{l,m}(\theta, \phi)$  with $m = l - a$, where $a = 0$ ($\blacksquare$), $1$ ($\times$) and $2$ ($\boxdot$), as a function of $l$  when $l$ goes from $a$ to $80$.}
\label{fig_fs3}
\end{figure}

According Eq. (\ref{fisher_shannon_complexity}), the Fisher-Shannon complexity of the three-dimensional rotator state $(l,m)$ is given by the Fisher-Shannon complexity of the density $\rho_{l,m}(\theta,\phi)$; that is,
\[
C_{FS}[\rho_{l,m}]=F[\rho_{l,m}]\times \frac{1}{2\pi e} e^{\frac{2}{3} S[\rho_{l,m}]}
= \left(4l(l+1)-2|m|(2l+1)\right)\times \frac{1}{2\pi e} e^{\frac{2}{3} S[\rho_{l,m}]},
\]
where the Shannon entropy $S[\rho_{l,m}]$ is given by Eq. (\ref{shannon_entropy}).  The variation of this complexity measure in terms of $l$ and $m$ is investigated in Figures \ref{fig_fs1}, \ref{fig_fs2} and \ref{fig_fs3}.
Figure \ref{fig_fs1} shows the values of the Fisher-Shannon complexity for fixed values of the angular quantum number $l=10,20,50,80$, for $m$ from $0$ to $l$. Notice that this complexity measure depends on the absolute value of $m$, so we have that $C_{FS}[\rho_{l,-m}] = C_{FS}[\rho_{l,m}]$.
In this case we observe that the function $C_{FS}[\rho_{l,m}]$ decreases monotonically as $m$ increases. We can also remark that the values of the complexity measure grow when $l$ increases.

Figure \ref{fig_fs2} shows specifically how the complexity measure grows with $l$ ($l\ge m$) for fixed values of $m$.

Finally, Figure \ref{fig_fs3} represents the values of $C_{FS}[\rho_{l,m}]$ as a function of $l$ when $m=l-a$ with $a=0,1,2$, for $l$ from $m$ to $80$. The complexity measure increases monotonically with $l$ in all the cases, and we see that
the larger the difference between $l$ and $m$, the higher the growth rate.

\subsection{Fisher-R\'enyi complexity}

\begin{figure}
\begin{center}
\includegraphics[width=10cm]{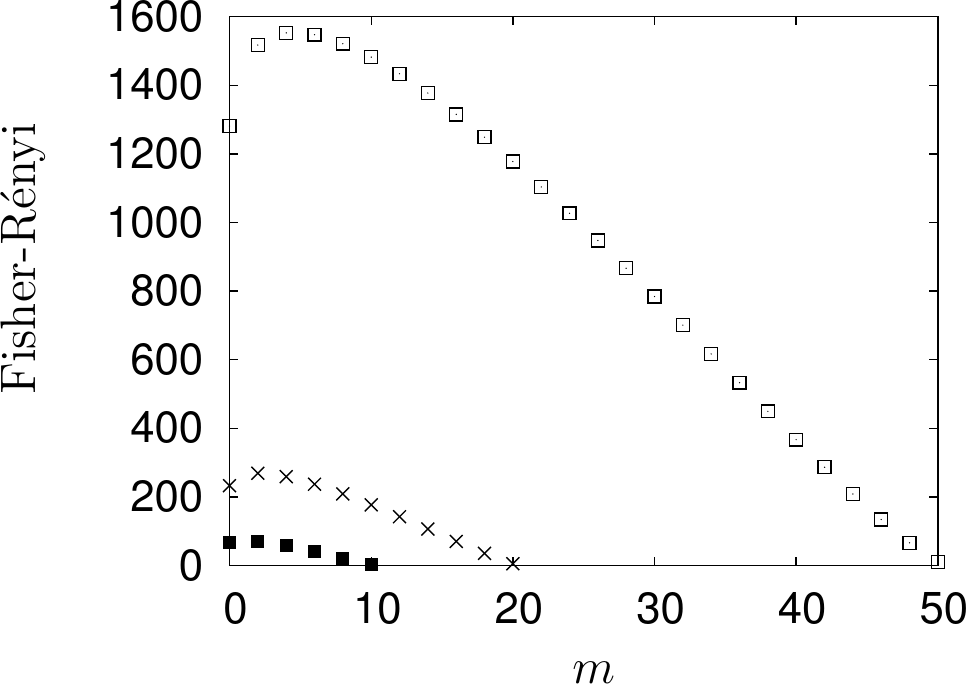}
\end{center}
\caption{Study of the  Fisher-R\'enyi complexity measure $C^{(2)}_{FR}$ in terms of $m$ for various spherical harmonics $Y_{l,m}(\theta, \phi)$  with fixed values of  $l = 10$ ($\blacksquare$), $20$ ($\times$) and $50$ ($\odot$).}
\label{fig_fr1}
\end{figure}

\begin{figure}
\begin{center}
\includegraphics[width=10cm]{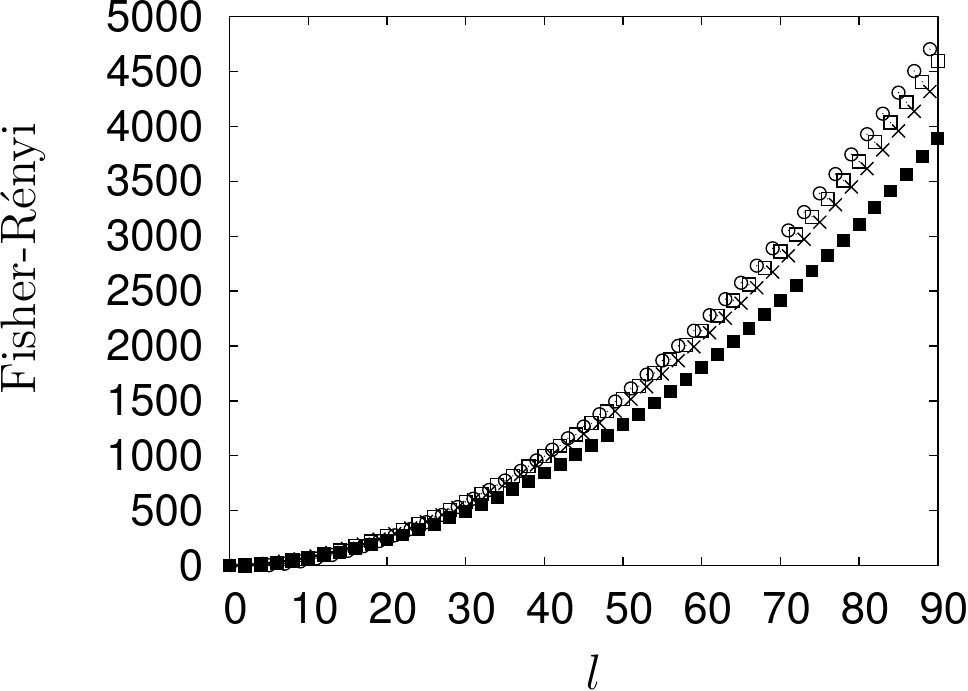}
\end{center}
\caption{Study of the  Fisher-R\'enyi complexity measure $C^{(2)}_{FR}$ in terms of $l$ for various spherical harmonics $Y_{l,m}(\theta, \phi)$  with fixed values of $m = 0$ ($\blacksquare$), $1$ ($\times$), $2$ ($\boxdot$) and $5$ ($\odot$).}
\label{fig_fr2}
\end{figure}

\begin{figure}
\begin{center}
\includegraphics[width=10cm]{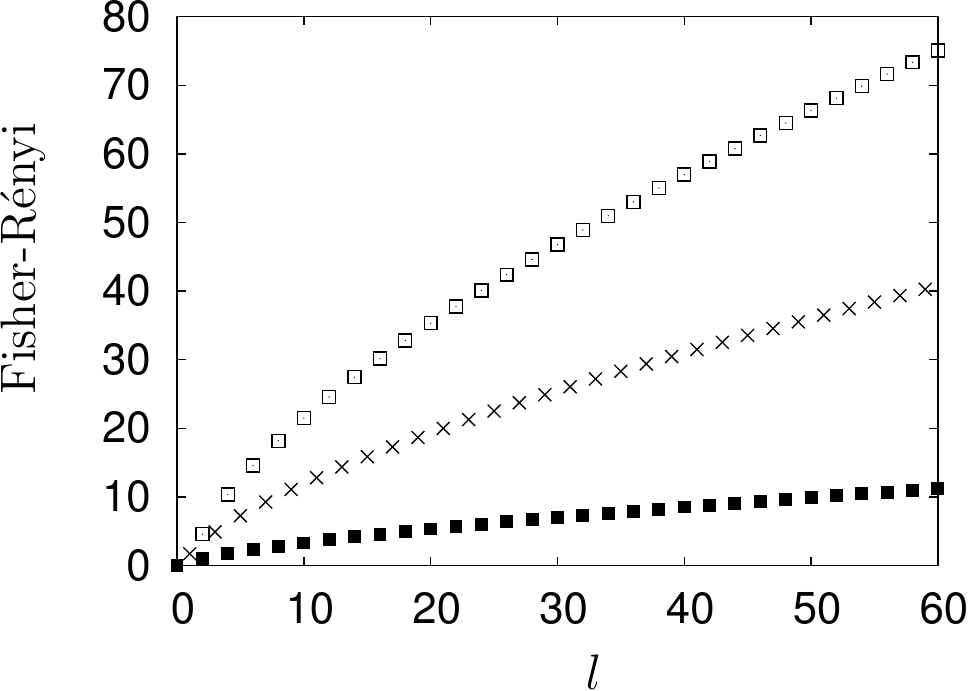}
\end{center}
\caption{Study of the  Fisher-R\'enyi complexity measure $C^{(2)}_{FR}$  for various spherical harmonics $Y_{l,m}(\theta, \phi)$  with $m = l - a$, where $a = 0$ ($\blacksquare$), $1$ ($\times$), and $2$ ($\odot$), as a function of $l$ when $l$ goes from $a$ to $60$.}
\label{fig_fr3}
\end{figure}

Following Eqs. (\ref{renyi_entropy}), (\ref{fisher_renyi}), (\ref{renyi_power_entropy}), (\ref{fisher_dimension_d3}) and (\ref{entropic_moment_harmonics}) we can express the Fisher-R\'enyi complexity $C_{FR}^{(q)}[\rho_{l,m}] $ of the three-dimensional rigid rotator in terms of the quantum numbers $l$ and $m$ via the entropic moments $W_q[\rho_{l,m}]$ already calculated in Section \ref{sec:information_measures} for any dimension; that is,
    \begin{eqnarray}
      \label{fisher_renyi_d}
      C_{FR}^{(q)}[\rho_{l,m}] 
        &=& \frac{1}{2 \pi e} F[\rho_{l,m}] \times W_q[\rho_{l,m}]^{\frac{2}{3(1-q)}} \nonumber\\
        &=& \frac{1}{2 \pi e} \left(4l(l+1)-2|m|(2l+1)\right) \times W_q[\rho_{l,m}]^{\frac{2}{3(1-q)}} ,\quad
        \text{with}\; q>0.
    \end{eqnarray}
Let us now explore the dependence of this complexity for a given $q$ (say e.g., $q = 2$) on the quantum parameters $l$ and $m$ by means of Figures \ref{fig_fr1}, \ref{fig_fr2} and  \ref{fig_fr3}.
Figure \ref{fig_fr1} represents  the Fisher-R\'enyi complexity measure $C_{FR}^{(q)}[\rho_{l,m}]$ for $q=2$ as a function of $m$ for fixed values of $l=10,20,50 $. The most notable feature of this figure is the maximum value achieved by this complexity measure for a given value $m_0\ge 0$ that depends on $l$ and $q$.
This contrasts with Figure \ref{fig_fs1}, where the maximum value of the Fisher-Shannon complexity measure is achieved for $m_0=0$ in all the cases.

Figure \ref{fig_fr2} shows the complexity $C_{FR}^{(q)}[\rho_{l,m}]$ for $q=2$ as a function of $l$ for $m=0,1,2,5$. We observe the same monotonically increasing behaviour shown by the Fisher-Shannon complexity in Figure \ref{fig_fs2}.

Figure \ref{fig_fr3} represents the complexity $C_{FR}^{(q)}[\rho_{l,m}]$ for $q=2$ as a function of $l$ for $m=l-a$, with $a=0,1,2$. This figure is completely analogous to the corresponding Figure \ref{fig_fs3} for the Fisher-Shannon complexity, where the complexity measure increases monotonically as $l$ grows.

\subsection{LMC complexity}

\begin{figure}
\begin{center}
\includegraphics[width=10cm]{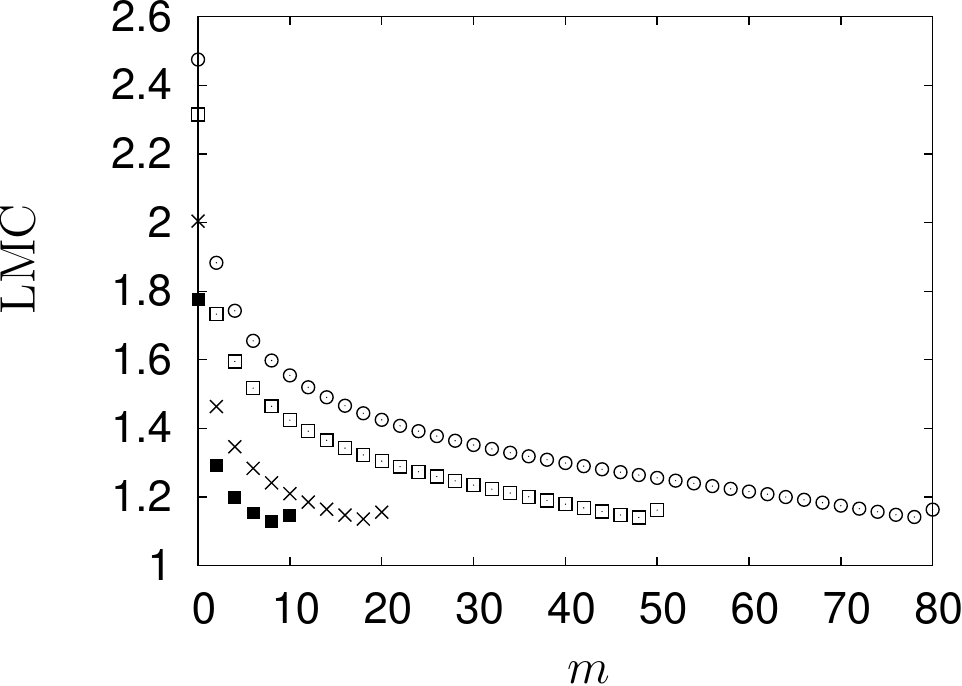}
\end{center}
\caption{Dependence of the LMC complexity measure on $m$ for various spherical harmonics $Y_{l,m}(\theta, \phi)$ with fixed orbital quantum number $l = 10$ ($\blacksquare$), $20$ ($\times$), $50$ ($\boxdot$), and $80$ ($\odot$).}
\label{fig_lmc1}
\end{figure}

According to Eqs. (\ref{shannon_entropy}), (\ref{lmc_complexity}) and (\ref{eq:def_disequilibrium}) we have that the LMC complexity of the rotator states $(l, m)$ is given by the expression
   \begin{equation}
      \label{lmc_d}
      C_{LMC}[\rho_{l,m}]=W_2[\rho_{l,m}] \times e^{S[\rho_{l,m}]}
   \end{equation}
where $W_2[\rho_{l,m}]$ have been already calculated in Section \ref{sec:information_measures}.
Figure \ref{fig_lmc1} shows the LMC complexity measure as a function of $m$ and fixed values $l=10,20,50,80$. This complexity measure has a decreasing behaviour as $m$ increases up to the position $m\sim l$ where a minimum is found and the complexity measure starts increasing.

\begin{figure}
\begin{center}
\includegraphics[width=10cm]{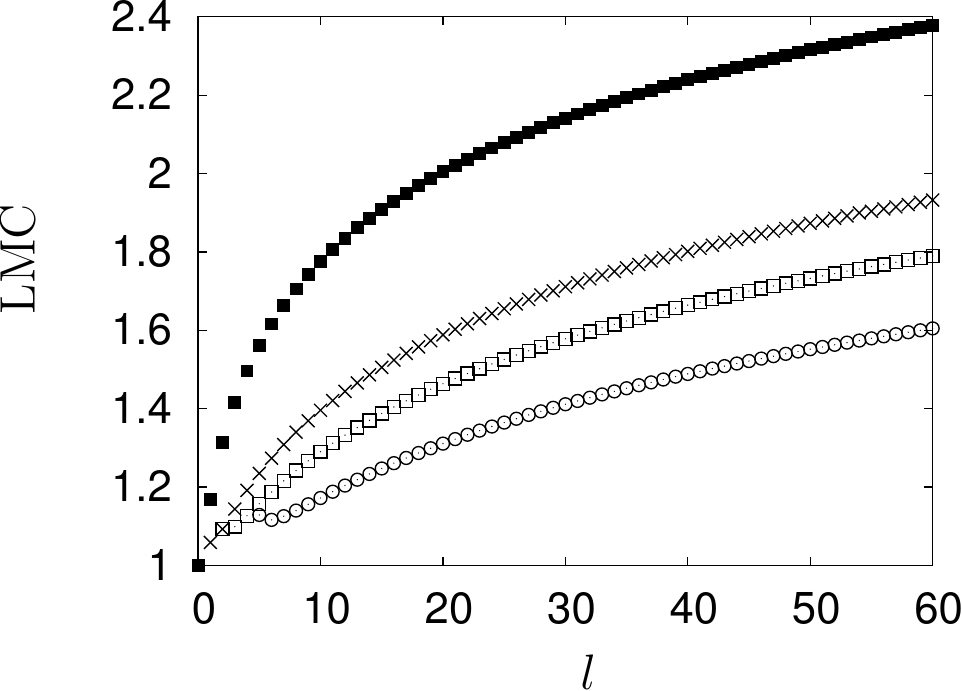}
\end{center}
\caption{Study of the LMC complexity measure as a function of $l$ for various spherical harmonics $Y_{l,m}(\theta, \phi)$ with fixed values of  $m = 0$ ($\blacksquare$), $1$ ($\times$), $2$ ($\boxdot$), and $5$ ($\odot$).}
\label{fig_lmc2}
\end{figure}

Figure \ref{fig_lmc2} shows the LMC complexity $C_{LMC}[\rho_{l,m}]$ as a function of $l$ for fixed values $m=0,1,2,5$. For $l\gg m$ this complexity have a clear increasing behaviour. But for some cases it has a minimum when $l\sim m$. These minima correspond to those found on Figure \ref{fig_lmc1}. They appear when the values of $l$ and $m$ have similar values.

\begin{figure}
\begin{center}
\includegraphics[width=10cm]{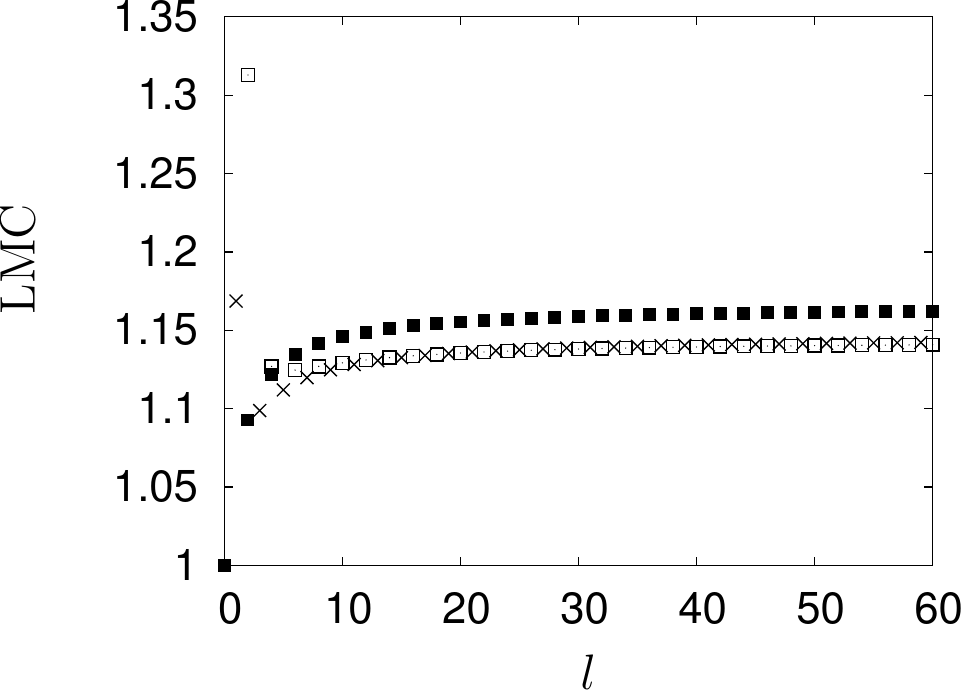}
\end{center}
\caption{Behavior of the LMC complexity measure for various spherical harmonics $Y_{l,m}(\theta, \phi)$  with $m = l-a$, where $a = 0$ ($\blacksquare$), $1$ ($\times$), and $2$ ($\boxdot$),  as a function of $l$ when $l$ goes from $a$ to $60$.}
\label{fig_lmc3}
\end{figure}

This behaviour is better explained in Figure \ref{fig_lmc3}, where $C_{LMC}[\rho_{l,m}]$ is represented as a function of $l$ for $m=l-a$, with $a=0,1,2$. Thus, $l\sim m$ in all the cases. We observe that for large  and moderate values of $l$ ($l\gtrsim 5$) the complexity measure is larger when $m=l$ than in the other two cases.

\section{Conclusions}

The rigid rotator model has been used in numerous mathematical and physical directions \cite{dai_13,avery_3,muller_1,atkinson12}; in particular it has been used to characterize the rotation of diatomic molecules (and is easily extended to linear polyatomic molecules), so that the entropy and complexity properties of these molecules can be referenced with respect to the corresponding rotator quantitites \cite{esquivel_pccp10}.  In this work we have investigated the entropy and complexity measures of the eigenfunctions of the $D$-dimensional rigid rotator model (namely, the hyperspherical functions) in terms of the dimensionality and the hyperquantum numbers which characterize them.

Since the hyperspherical harmonics describe the angular part of the stationary states of any central potential with arbitrary dimensionality, these information-theoretic quantities provide estimations for the angular anisotropy of the eigenfunctions of a central potential in the hyperspace. In other terms, they quantify the rich variety of $D$-dimensional geometries of the lobe-structure of the quantum states of the corresponding system (e.g., hydrogenic orbitals for the hydrogen atom), which are described by means of $D$ integer hypernumbers (e.g., the principal, orbital and azimuthal quantum numbers $n$, $l$ and $m$, in the three dimensional case).

Specifically, besides the explicit expression for the Fisher information, first we have found
the entropic or frequency moments of the hyperspherical harmonics, which allows one to find the R\'enyi and Tsallis entropies of the rigid rotator in a straightforward manner.
Then, we numerically study the dependence on the quantum numbers $(l, m)$ for the complexity measures of Fisher-Shannon, Fisher-R\'enyi and LMC types of the spherical harmonics $Y_{l,m}(\theta,\phi)$, which are the eigenfunctions of the three-dimensional rigid rotator.

Let us highlight that the spatial complexity of the associated probability densities (\ref{eq:3d_density}) to the spherical harmonics is clearly related to the number of lobes of their three-dimensional representations. In fact the degree of the involved Gegenbauer polynomial is connected to its number of maxima, and hence to the number of lobes, that is equal to $l - |m| + 1$; so the complexity is expected to grow as the difference $l - |m|$ increases. This behaviour is only grasped by the Fisher-Shannon complexity. Indeed the Fisher-R\'enyi and the LMC complexities, although follow this behaviour in most cases, show pointwise differences with respect to the Fisher-Shannon complexity. This can be seen e.g. in Figure \ref{fig_fs1}, where the Fisher-R\'enyi measure increases with $|m|$ at low values of $|m|$. Similarly, this counterintuitive behaviour can also be seen for the LMC complexity in Figures \ref{fig_lmc1} and \ref{fig_lmc2} for the cases where $l \simeq |m|$. As well, this phenomena is also apparent in Figure \ref{fig_lmc3} in a transparent manner, where $|m| \simeq l$ in all the cases and a clear monotonic behaviour in the plotted data is not observed. In turn, it is remarkable that the Fisher-Shannon complexity grasps the visual, intuitive complexity of the  density associated to the spherical harmonics. From this point of view we can endorse this quantity as the most appropriate complexity measure in this system.

Finally, let us also point out that the entropy and complexity quantities used in this work do not only quantify the anisotropic character of the stationary states of the central potentials in any dimensionality, but they can potentially be used to visualize $D$-dimensional models that are becoming integral components of data processing in many fields, including medicine, chemistry, architecture, agriculture and biology over last few years. Moreover, they could be employed to carry out volumetric shape analyses which permit an evaluation of the actual structures that are implicitly represented in $D$-dimensional image data.

\section*{Acknowledgements}

This work has been partially funded by the Junta-de-Andalucía grants FQM-207, FQM-7276 and FQM-4643, as well as the MICINN grant FIS2011-24540


\begin{thebibliography}{46}
\expandafter\ifx\csname natexlab\endcsname\relax\def\natexlab#1{#1}\fi
\expandafter\ifx\csname bibnamefont\endcsname\relax
  \def\bibnamefont#1{#1}\fi
\expandafter\ifx\csname bibfnamefont\endcsname\relax
  \def\bibfnamefont#1{#1}\fi
\expandafter\ifx\csname citenamefont\endcsname\relax
  \def\citenamefont#1{#1}\fi
\expandafter\ifx\csname url\endcsname\relax
  \def\url#1{\texttt{#1}}\fi
\expandafter\ifx\csname urlprefix\endcsname\relax\def\urlprefix{URL }\fi
\providecommand{\bibinfo}[2]{#2}
\providecommand{\eprint}[2][]{\url{#2}}

\bibitem[{\citenamefont{Galindo and Pascual}(1990)}]{galindo_1}
\bibinfo{author}{\bibfnamefont{A.}~\bibnamefont{Galindo}} \bibnamefont{and}
  \bibinfo{author}{\bibfnamefont{P.}~\bibnamefont{Pascual}},
  \emph{\bibinfo{title}{Quantum {M}echanics}} (\bibinfo{publisher}{Springer,
  Berlin}, \bibinfo{year}{1990}).

\bibitem[{\citenamefont{Herschbach et~al.}(1993)\citenamefont{Herschbach,
  Avery, and Goscinski}}]{herschbach_1}
\bibinfo{author}{\bibfnamefont{D.~R.} \bibnamefont{Herschbach}},
  \bibinfo{author}{\bibfnamefont{J.}~\bibnamefont{Avery}}, \bibnamefont{and}
  \bibinfo{author}{\bibfnamefont{O.}~\bibnamefont{Goscinski}},
  \emph{\bibinfo{title}{Dimensional {S}caling in {C}hemical {P}hysics}}
  (\bibinfo{publisher}{Kluwer, Dordrecht}, \bibinfo{year}{1993}).

\bibitem[{\citenamefont{Bouvrie et~al.}(2011)\citenamefont{Bouvrie, Angulo, and
  Dehesa}}]{bouvrie_1}
\bibinfo{author}{\bibfnamefont{P.~A.} \bibnamefont{Bouvrie}},
  \bibinfo{author}{\bibfnamefont{J.~C.} \bibnamefont{Angulo}},
  \bibnamefont{and} \bibinfo{author}{\bibfnamefont{J.~S.}
  \bibnamefont{Dehesa}}, \bibinfo{journal}{Physica A}
  \textbf{\bibinfo{volume}{390}}, \bibinfo{pages}{2215} (\bibinfo{year}{2011}).

\bibitem[{\citenamefont{Dehesa et~al.}(2010)\citenamefont{Dehesa, L\'opez-Rosa,
  Mart\'inez-Finkelshtein, and Y\'a\~nez}}]{dehesa_3}
\bibinfo{author}{\bibfnamefont{J.~S.} \bibnamefont{Dehesa}},
  \bibinfo{author}{\bibfnamefont{S.}~\bibnamefont{L\'opez-Rosa}},
  \bibinfo{author}{\bibfnamefont{A.}~\bibnamefont{Mart\'inez-Finkelshtein}},
  \bibnamefont{and} \bibinfo{author}{\bibfnamefont{R.~J.}
  \bibnamefont{Y\'a\~nez}}, \bibinfo{journal}{Int. J. Quantum Chemistry}
  \textbf{\bibinfo{volume}{109}}, \bibinfo{pages}{1529} (\bibinfo{year}{2010}).

\bibitem[{\citenamefont{Dehesa et~al.}(1994)\citenamefont{Dehesa, Van~Assche,
  and Y\'a\~nez}}]{dehesa_5}
\bibinfo{author}{\bibfnamefont{J.~S.} \bibnamefont{Dehesa}},
  \bibinfo{author}{\bibfnamefont{W.}~\bibnamefont{Van~Assche}},
  \bibnamefont{and} \bibinfo{author}{\bibfnamefont{R.~J.}
  \bibnamefont{Y\'a\~nez}}, \bibinfo{journal}{Phys. Rev. A}
  \textbf{\bibinfo{volume}{50}}, \bibinfo{pages}{3065} (\bibinfo{year}{1994}).

\bibitem[{\citenamefont{Dehesa et~al.}(1998)\citenamefont{Dehesa, Y\'a{\~n}ez,
  I., and V.}}]{dehesa_6}
\bibinfo{author}{\bibfnamefont{J.~S.} \bibnamefont{Dehesa}},
  \bibinfo{author}{\bibfnamefont{R.~J.} \bibnamefont{Y\'a{\~n}ez}},
  \bibinfo{author}{\bibfnamefont{A.~A.} \bibnamefont{I.}}, \bibnamefont{and}
  \bibinfo{author}{\bibfnamefont{B.}~\bibnamefont{V.}}, \bibinfo{journal}{J.
  Math. Phys.} \textbf{\bibinfo{volume}{39}}, \bibinfo{pages}{3050}
  (\bibinfo{year}{1998}).

\bibitem[{\citenamefont{L\'opez-Rosa et~al.}(2011)\citenamefont{L\'opez-Rosa,
  Montero, S\'anchez-Moreno, Venegas, and Dehesa}}]{lopez_1}
\bibinfo{author}{\bibfnamefont{S.}~\bibnamefont{L\'opez-Rosa}},
  \bibinfo{author}{\bibfnamefont{J.}~\bibnamefont{Montero}},
  \bibinfo{author}{\bibfnamefont{P.}~\bibnamefont{S\'anchez-Moreno}},
  \bibinfo{author}{\bibfnamefont{J.}~\bibnamefont{Venegas}}, \bibnamefont{and}
  \bibinfo{author}{\bibfnamefont{J.}~\bibnamefont{Dehesa}},
  \bibinfo{journal}{J. Math. Chem.} \textbf{\bibinfo{volume}{49}},
  \bibinfo{pages}{971} (\bibinfo{year}{2011}).

\bibitem[{\citenamefont{Dehesa et~al.}(2011)\citenamefont{Dehesa, L\'opez-Rosa,
  and Manzano}}]{dehesa_4}
\bibinfo{author}{\bibfnamefont{J.~S.} \bibnamefont{Dehesa}},
  \bibinfo{author}{\bibfnamefont{S.}~\bibnamefont{L\'opez-Rosa}},
  \bibnamefont{and} \bibinfo{author}{\bibfnamefont{D.}~\bibnamefont{Manzano}},
  \emph{\bibinfo{title}{{E}ntropy and complexity analyses of {D}-dimensional
  quantum systems. {I}n {S}tatistical {C}omplexity: Applications in
  {E}lectronic {S}tructure, edited by {K}. {D}. {S}en}}
  (\bibinfo{publisher}{Springer Verlag, Berlin}, \bibinfo{year}{2011}).

\bibitem[{\citenamefont{Dehesa et~al.}(2001)\citenamefont{Dehesa,
  Mart\'inez-Finkelshtein, and S\'anchez-Ruiz}}]{dehesa_7}
\bibinfo{author}{\bibfnamefont{J.~S.} \bibnamefont{Dehesa}},
  \bibinfo{author}{\bibfnamefont{A.}~\bibnamefont{Mart\'inez-Finkelshtein}},
  \bibnamefont{and}
  \bibinfo{author}{\bibfnamefont{J.}~\bibnamefont{S\'anchez-Ruiz}},
  \bibinfo{journal}{J. Comput. Appl. Math.} \textbf{\bibinfo{volume}{133}},
  \bibinfo{pages}{23} (\bibinfo{year}{2001}).

\bibitem[{\citenamefont{Curilef and Tsallis}(1995)}]{curilef_1}
\bibinfo{author}{\bibfnamefont{S.}~\bibnamefont{Curilef}} \bibnamefont{and}
  \bibinfo{author}{\bibfnamefont{C.}~\bibnamefont{Tsallis}},
  \bibinfo{journal}{Physica A} \textbf{\bibinfo{volume}{215}},
  \bibinfo{pages}{542} (\bibinfo{year}{1995}).

\bibitem[{\citenamefont{Rist and Faure}(2012)}]{rist_jmc12}
\bibinfo{author}{\bibfnamefont{C.}~\bibnamefont{Rist}} \bibnamefont{and}
  \bibinfo{author}{\bibfnamefont{A.}~\bibnamefont{Faure}}, \bibinfo{journal}{J.
  Math. Chem.} \textbf{\bibinfo{volume}{50}}, \bibinfo{pages}{588}
  (\bibinfo{year}{2012}).

\bibitem[{\citenamefont{Gartner et~al.}(2013)\citenamefont{Gartner, Omiste,
  Schmelcher, and Gonz\'alez-F\'erez}}]{gartner_arxiv13}
\bibinfo{author}{\bibfnamefont{M.}~\bibnamefont{Gartner}},
  \bibinfo{author}{\bibfnamefont{J.~J.} \bibnamefont{Omiste}},
  \bibinfo{author}{\bibfnamefont{P.}~\bibnamefont{Schmelcher}},
  \bibnamefont{and}
  \bibinfo{author}{\bibfnamefont{R.}~\bibnamefont{Gonz\'alez-F\'erez}},
  \bibinfo{journal}{Molecular Phys.} \textbf{\bibinfo{volume}{111}},
  \bibinfo{pages}{1865} (\bibinfo{year}{2013}).

\bibitem[{\citenamefont{Dai and Xu}(2013)}]{dai_13}
\bibinfo{author}{\bibfnamefont{F.}~\bibnamefont{Dai}} \bibnamefont{and}
  \bibinfo{author}{\bibfnamefont{Y.}~\bibnamefont{Xu}},
  \emph{\bibinfo{title}{Approximation Theory and Harmonic Analysis on Spheres
  and Balls}} (\bibinfo{publisher}{Springer, Berlin}, \bibinfo{year}{2013}).

\bibitem[{\citenamefont{M\"uller}(1997)}]{muller_1}
\bibinfo{author}{\bibfnamefont{C.}~\bibnamefont{M\"uller}},
  \emph{\bibinfo{title}{Analysis of Spherical Harmonics in Euclidean Space}}
  (\bibinfo{publisher}{Springer, Berlin}, \bibinfo{year}{1997}).

\bibitem[{\citenamefont{Atkinson and Han}(2012)}]{atkinson12}
\bibinfo{author}{\bibfnamefont{K.}~\bibnamefont{Atkinson}} \bibnamefont{and}
  \bibinfo{author}{\bibfnamefont{W.}~\bibnamefont{Han}}, in
  \emph{\bibinfo{booktitle}{Lecture Notes in Mathematics}}
  (\bibinfo{publisher}{Springer}, \bibinfo{year}{2012}), vol.
  \bibinfo{volume}{124}.

\bibitem[{\citenamefont{Avery}(1989)}]{avery_3}
\bibinfo{author}{\bibfnamefont{J.}~\bibnamefont{Avery}},
  \emph{\bibinfo{title}{Hyperspherical {H}armonics: {A}pplications in {Q}uantum
  {T}heory}} (\bibinfo{publisher}{Kluwer, Dordrecht}, \bibinfo{year}{1989}).

\bibitem[{\citenamefont{Avery}(2000)}]{avery_2}
\bibinfo{author}{\bibfnamefont{J.}~\bibnamefont{Avery}},
  \emph{\bibinfo{title}{{H}yperspherical {H}armonics and {G}eneralized
  {S}turmians}} (\bibinfo{publisher}{Kluwer, Dordrecht}, \bibinfo{year}{2000}).

\bibitem[{\citenamefont{Kyriakopoulos}(1968)}]{kyriakopoulos_1}
\bibinfo{author}{\bibfnamefont{E.}~\bibnamefont{Kyriakopoulos}},
  \bibinfo{journal}{Phys. Rev.} \textbf{\bibinfo{volume}{174}},
  \bibinfo{pages}{1846} (\bibinfo{year}{1968}).

\bibitem[{\citenamefont{Stein and Weiss}(1971)}]{stein_1}
\bibinfo{author}{\bibfnamefont{E.~M.} \bibnamefont{Stein}} \bibnamefont{and}
  \bibinfo{author}{\bibfnamefont{G.}~\bibnamefont{Weiss}},
  \emph{\bibinfo{title}{Fourier {A}nalysis in {E}ucledian {S}paces}}
  (\bibinfo{publisher}{Princeton University Press, Princeton, New Jersey},
  \bibinfo{year}{1971}).

\bibitem[{\citenamefont{Aquilanti et~al.}(1986)\citenamefont{Aquilanti,
  Cavalli, and Grossi}}]{aquilanti_1}
\bibinfo{author}{\bibfnamefont{V.}~\bibnamefont{Aquilanti}},
  \bibinfo{author}{\bibfnamefont{S.}~\bibnamefont{Cavalli}}, \bibnamefont{and}
  \bibinfo{author}{\bibfnamefont{G.}~\bibnamefont{Grossi}},
  \bibinfo{journal}{J. Chem. Phys.} \textbf{\bibinfo{volume}{85}},
  \bibinfo{pages}{1362} (\bibinfo{year}{1986}).

\bibitem[{\citenamefont{Nikiforov et~al.}(1991)\citenamefont{Nikiforov, Suslov,
  and Uvarov}}]{nikiforov_2}
\bibinfo{author}{\bibfnamefont{A.~F.} \bibnamefont{Nikiforov}},
  \bibinfo{author}{\bibfnamefont{S.~K.} \bibnamefont{Suslov}},
  \bibnamefont{and} \bibinfo{author}{\bibfnamefont{V.~B.}
  \bibnamefont{Uvarov}}, \emph{\bibinfo{title}{{C}lassical {O}rthogonal
  {P}olynomials of a {D}iscrete {V}ariable}} (\bibinfo{publisher}{Springer
  Verlag, Berlin}, \bibinfo{year}{1991}).

\bibitem[{\citenamefont{Avery}(1993)}]{avery_4}
\bibinfo{author}{\bibfnamefont{J.}~\bibnamefont{Avery}}, \bibinfo{journal}{J.
  Phys. Chem.} \textbf{\bibinfo{volume}{97}}, \bibinfo{pages}{2406}
  (\bibinfo{year}{1993}).

\bibitem[{\citenamefont{Wen and Avery}(1985)}]{wen_1}
\bibinfo{author}{\bibfnamefont{Z.~Y.} \bibnamefont{Wen}} \bibnamefont{and}
  \bibinfo{author}{\bibfnamefont{J.}~\bibnamefont{Avery}}, \bibinfo{journal}{J.
  Math. Phys.} \textbf{\bibinfo{volume}{26}}, \bibinfo{pages}{396}
  (\bibinfo{year}{1985}).

\bibitem[{\citenamefont{Avery}(1998)}]{avery_1}
\bibinfo{author}{\bibfnamefont{J.}~\bibnamefont{Avery}}, \bibinfo{journal}{J.
  Math. Chem.} \textbf{\bibinfo{volume}{24}}, \bibinfo{pages}{169}
  (\bibinfo{year}{1998}).

\bibitem[{\citenamefont{Dehesa et~al.}(2007)\citenamefont{Dehesa, L\'opez-Rosa,
  and Y\'a\~nez}}]{dehesa_2}
\bibinfo{author}{\bibfnamefont{J.~S.} \bibnamefont{Dehesa}},
  \bibinfo{author}{\bibfnamefont{S.}~\bibnamefont{L\'opez-Rosa}},
  \bibnamefont{and} \bibinfo{author}{\bibfnamefont{R.~J.}
  \bibnamefont{Y\'a\~nez}}, \bibinfo{journal}{J. Math. Phys.}
  \textbf{\bibinfo{volume}{48}}, \bibinfo{pages}{043503}
  (\bibinfo{year}{2007}).

\bibitem[{\citenamefont{Uffink}(1990)}]{uffink_1}
\bibinfo{author}{\bibfnamefont{J.}~\bibnamefont{Uffink}},
  \emph{\bibinfo{title}{Measures of {U}ncertainty and the {U}ncertainty
  {P}rinciple}} (\bibinfo{publisher}{PhD Thesis, University of Utrecht},
  \bibinfo{year}{1990}), \bibinfo{note}{see also references herein}.

\bibitem[{\citenamefont{Parr and Yang}(1989)}]{parr_1}
\bibinfo{author}{\bibfnamefont{R.~G.} \bibnamefont{Parr}} \bibnamefont{and}
  \bibinfo{author}{\bibfnamefont{W.}~\bibnamefont{Yang}},
  \emph{\bibinfo{title}{Density {F}unctional {T}heory of {A}toms and
  {M}olecules}} (\bibinfo{publisher}{Oxford University Press, Oxford},
  \bibinfo{year}{1989}).

\bibitem[{\citenamefont{Liu and Parr}(1996)}]{lin_1}
\bibinfo{author}{\bibfnamefont{S.}~\bibnamefont{Liu}} \bibnamefont{and}
  \bibinfo{author}{\bibfnamefont{R.~G.} \bibnamefont{Parr}},
  \bibinfo{journal}{Phys. Rev. A} \textbf{\bibinfo{volume}{53}},
  \bibinfo{pages}{2211} (\bibinfo{year}{1996}).

\bibitem[{\citenamefont{Liu and Parr}(1997)}]{lin_2}
\bibinfo{author}{\bibfnamefont{S.}~\bibnamefont{Liu}} \bibnamefont{and}
  \bibinfo{author}{\bibfnamefont{R.~G.} \bibnamefont{Parr}},
  \bibinfo{journal}{Physica A} \textbf{\bibinfo{volume}{55}},
  \bibinfo{pages}{1792} (\bibinfo{year}{1997}).

\bibitem[{\citenamefont{Nagy et~al.}(1999)\citenamefont{Nagy, Liu, and
  Parr}}]{nagy_1}
\bibinfo{author}{\bibfnamefont{A.}~\bibnamefont{Nagy}},
  \bibinfo{author}{\bibfnamefont{S.}~\bibnamefont{Liu}}, \bibnamefont{and}
  \bibinfo{author}{\bibfnamefont{R.~G.} \bibnamefont{Parr}},
  \bibinfo{journal}{Phys. Rev. A} \textbf{\bibinfo{volume}{59}},
  \bibinfo{pages}{3349} (\bibinfo{year}{1999}).

\bibitem[{\citenamefont{Angulo et~al.}(2000)\citenamefont{Angulo, Romera, and
  Dehesa}}]{angulo_2}
\bibinfo{author}{\bibfnamefont{J.~C.} \bibnamefont{Angulo}},
  \bibinfo{author}{\bibfnamefont{E.}~\bibnamefont{Romera}}, \bibnamefont{and}
  \bibinfo{author}{\bibfnamefont{J.~S.} \bibnamefont{Dehesa}},
  \bibinfo{journal}{J. Math. Phys.} \textbf{\bibinfo{volume}{41}},
  \bibinfo{pages}{7906} (\bibinfo{year}{2000}).

\bibitem[{\citenamefont{Pintarelli and Vericat}(2003)}]{pinta_1}
\bibinfo{author}{\bibfnamefont{M.~B.} \bibnamefont{Pintarelli}}
  \bibnamefont{and} \bibinfo{author}{\bibfnamefont{F.}~\bibnamefont{Vericat}},
  \bibinfo{journal}{Physica A} \textbf{\bibinfo{volume}{324}},
  \bibinfo{pages}{568} (\bibinfo{year}{2003}).

\bibitem[{\citenamefont{Romera et~al.}(2001)\citenamefont{Romera, Angulo, and
  Dehesa}}]{romera_1}
\bibinfo{author}{\bibfnamefont{E.}~\bibnamefont{Romera}},
  \bibinfo{author}{\bibfnamefont{J.~C.} \bibnamefont{Angulo}},
  \bibnamefont{and} \bibinfo{author}{\bibfnamefont{J.~S.}
  \bibnamefont{Dehesa}}, \bibinfo{journal}{J. Math. Phys.}
  \textbf{\bibinfo{volume}{42}}, \bibinfo{pages}{2309} (\bibinfo{year}{2001}).

\bibitem[{\citenamefont{Leonenko et~al.}(2008)\citenamefont{Leonenko, Pronzato,
  and Savani}}]{leo_1}
\bibinfo{author}{\bibfnamefont{N.}~\bibnamefont{Leonenko}},
  \bibinfo{author}{\bibfnamefont{L.}~\bibnamefont{Pronzato}}, \bibnamefont{and}
  \bibinfo{author}{\bibfnamefont{V.}~\bibnamefont{Savani}},
  \bibinfo{journal}{Ann. Stat.} \textbf{\bibinfo{volume}{40}},
  \bibinfo{pages}{2153} (\bibinfo{year}{2008}).

\bibitem[{\citenamefont{Dette et~al.}(2005)\citenamefont{Dette, Melas, and
  Pepelyshev}}]{dette_as05}
\bibinfo{author}{\bibfnamefont{H.}~\bibnamefont{Dette}},
  \bibinfo{author}{\bibfnamefont{V.~B.} \bibnamefont{Melas}}, \bibnamefont{and}
  \bibinfo{author}{\bibfnamefont{A.}~\bibnamefont{Pepelyshev}},
  \bibinfo{journal}{Ann. Stat.} \textbf{\bibinfo{volume}{33}},
  \bibinfo{pages}{2758} (\bibinfo{year}{2005}).

\bibitem[{\citenamefont{R\'enyi}(1970)}]{renyi_1}
\bibinfo{author}{\bibfnamefont{A.}~\bibnamefont{R\'enyi}},
  \emph{\bibinfo{title}{Probability {T}heory}} (\bibinfo{publisher}{Akademi
  Kiado, Budapest}, \bibinfo{year}{1970}).

\bibitem[{\citenamefont{Tsallis}(1998)}]{tsallis_jsp88}
\bibinfo{author}{\bibfnamefont{C.}~\bibnamefont{Tsallis}}, \bibinfo{journal}{J.
  Stat. Phys.} \textbf{\bibinfo{volume}{52}}, \bibinfo{pages}{479}
  (\bibinfo{year}{1998}).

\bibitem[{\citenamefont{Fisher}(1925)}]{fisher_1}
\bibinfo{author}{\bibfnamefont{R.~A.} \bibnamefont{Fisher}},
  \bibinfo{journal}{Proc. Cambridge Philos. Soc.}
  \textbf{\bibinfo{volume}{22}}, \bibinfo{pages}{700} (\bibinfo{year}{1925}).

\bibitem[{\citenamefont{Frieden}(2004)}]{frieden_1}
\bibinfo{author}{\bibfnamefont{B.~R.} \bibnamefont{Frieden}},
  \emph{\bibinfo{title}{Science from {F}isher Information}}
  (\bibinfo{publisher}{Cambridge University Press, Cambridge},
  \bibinfo{year}{2004}).

\bibitem[{\citenamefont{Sears and y~otros}(1980)}]{sears_1}
\bibinfo{author}{\bibfnamefont{S.~B.} \bibnamefont{Sears}} \bibnamefont{and}
  \bibinfo{author}{\bibfnamefont{R.~G.~P.} \bibnamefont{y~otros}},
  \bibinfo{journal}{Israel J. Chem.} \textbf{\bibinfo{volume}{19}},
  \bibinfo{pages}{165} (\bibinfo{year}{1980}).

\bibitem[{\citenamefont{Romera and Nagy}(2008)}]{romera_2}
\bibinfo{author}{\bibfnamefont{E.}~\bibnamefont{Romera}} \bibnamefont{and}
  \bibinfo{author}{\bibfnamefont{A.}~\bibnamefont{Nagy}},
  \bibinfo{journal}{Phys. Lett. A} \textbf{\bibinfo{volume}{372}},
  \bibinfo{pages}{6823} (\bibinfo{year}{2008}).

\bibitem[{\citenamefont{Catalan et~al.}(2002)\citenamefont{Catalan, Garay, and
  L\'opez-Ruiz}}]{catalan_2}
\bibinfo{author}{\bibfnamefont{R.~G.} \bibnamefont{Catalan}},
  \bibinfo{author}{\bibfnamefont{J.}~\bibnamefont{Garay}}, \bibnamefont{and}
  \bibinfo{author}{\bibfnamefont{R.}~\bibnamefont{L\'opez-Ruiz}},
  \bibinfo{journal}{Phys. Rev. E} \textbf{\bibinfo{volume}{66}},
  \bibinfo{pages}{011102} (\bibinfo{year}{2002}).

\bibitem[{\citenamefont{Ru\'iz}(2005)}]{lopez_ruiz_2}
\bibinfo{author}{\bibfnamefont{R.~L.} \bibnamefont{Ru\'iz}},
  \bibinfo{journal}{Biophys. Chem.} \textbf{\bibinfo{volume}{115}},
  \bibinfo{pages}{215} (\bibinfo{year}{2005}).

\bibitem[{\citenamefont{Romera et~al.}(2006)\citenamefont{Romera,
  S\'anchez-Moreno, and Dehesa}}]{romera_jmp06}
\bibinfo{author}{\bibfnamefont{E.}~\bibnamefont{Romera}},
  \bibinfo{author}{\bibfnamefont{P.}~\bibnamefont{S\'anchez-Moreno}},
  \bibnamefont{and} \bibinfo{author}{\bibfnamefont{J.~S.}
  \bibnamefont{Dehesa}}, \bibinfo{journal}{J. Math. Phys.}
  \textbf{\bibinfo{volume}{47}}, \bibinfo{pages}{103504}
  (\bibinfo{year}{2006}).

\bibitem[{\citenamefont{Srivastava}(1988)}]{srivastava_ass88}
\bibinfo{author}{\bibfnamefont{H.~M.} \bibnamefont{Srivastava}},
  \bibinfo{journal}{Astrophys. Space Science} \textbf{\bibinfo{volume}{150}},
  \bibinfo{pages}{251} (\bibinfo{year}{1988}).

\bibitem[{\citenamefont{Esquivel et~al.}(2010)\citenamefont{Esquivel, C.Angulo,
  Antol\'in, Dehesa, L\'opez-Rosa, and Flores-Gallegos}}]{esquivel_pccp10}
\bibinfo{author}{\bibfnamefont{R.~O.} \bibnamefont{Esquivel}},
  \bibinfo{author}{\bibfnamefont{J.}~\bibnamefont{C.Angulo}},
  \bibinfo{author}{\bibfnamefont{J.}~\bibnamefont{Antol\'in}},
  \bibinfo{author}{\bibfnamefont{J.~S.} \bibnamefont{Dehesa}},
  \bibinfo{author}{\bibfnamefont{S.}~\bibnamefont{L\'opez-Rosa}},
  \bibnamefont{and}
  \bibinfo{author}{\bibfnamefont{N.}~\bibnamefont{Flores-Gallegos}},
  \bibinfo{journal}{Phys. Chem. Chem. Phys.} \textbf{\bibinfo{volume}{12}},
  \bibinfo{pages}{7108} (\bibinfo{year}{2010}).

\end{thebibliography}

\end{document}